\input epsf

\magnification\magstephalf
\overfullrule 0pt
\def\gsim{\raise.3ex\hbox{$\;>$\kern-.75em\lower1ex\hbox{$\sim$}$\;$}}

\font\rfont=cmr10 at 10 true pt
\def\ref#1{$^{\hbox{\rfont {[#1]}}}$}


\font\fourteenbf=cmbx12 scaled\magstep1

\font\tenbfit=cmbxti10
\font\sevenbfit=cmbxti10 at 7pt
\font\fivebfit=cmbxti10 at 5pt
\newfam\bfitfam 
\textfont\bfitfam=\tenbfit  \scriptfont\bfitfam=\sevenbfit
\scriptscriptfont\bfitfam=\fivebfit

\font\tenbfit=cmbxti10
\font\sevenbfit=cmbxti10 at 7pt
\font\fivebfit=cmbxti10 at 5pt
\newfam\bfitfam 
\textfont\bfitfam=\tenbfit  \scriptfont\bfitfam=\sevenbfit
\scriptscriptfont\bfitfam=\fivebfit

\font\tenbit=cmmib10
\newfam\bitfam
\textfont\bitfam=\tenbit%

\font\tenmbf=cmbx10
\font\sevenmbf=cmbx7
\font\fivembf=cmbx5
\newfam\mbffam
\textfont\mbffam=\tenmbf \scriptfont\mbffam=\sevenmbf
\scriptscriptfont\mbffam=\fivembf

\font\tenbsy=cmbsy10
\newfam\bsyfam 
\textfont\bsyfam=\tenbsy%


\def\pmb#1{\setbox0=\hbox{#1}
 \kern.05em\copy0\kern-\wd0 \kern-.025em\raise.0433em\box0 }

\def\slash{/\kern-.5em}


 %


\def\boxit#1{\vbox{\hrule\hbox{\vrule\kern1pt\vbox
{\kern1pt#1\kern1pt}\kern1pt\vrule}\hrule}}

\def\h{\hfill\break}
\parskip=6pt
\parindent=0pt
\hsize=17truecm\hoffset=-5truemm
\vsize=23truecm
\def\footnoterule{\kern-3pt
\hrule width 17truecm \kern 2.6pt}


\catcode`\@=11 

\def\nolabels{\def\wrlabeL##1{}\def\eqlabeL##1{}\def\reflabeL##1{}}
\def\writelabels{\def\wrlabeL##1{\leavevmode\vadjust{\rlap{\smash%
{\line{{\escapechar=` \hfill\rlap{\sevenrm\hskip.03in\string##1}}}}}}}%
\def\eqlabeL##1{{\escapechar-1\rlap{\sevenrm\hskip.05in\string##1}}}%
\def\reflabeL##1{\noexpand\llap{\noexpand\sevenrm\string\string\string##1}}}
\nolabels
\global\newcount\refno \global\refno=1
\newwrite\rfile
\def\defref{$^{{\hbox{\rfont [\the\refno]}}}$\nref}
\def\nref#1{\xdef#1{\the\refno}\writedef{#1\leftbracket#1}%
\ifnum\refno=1\immediate\openout\rfile=refs.tmp\fi
\global\advance\refno by1\chardef\wfile=\rfile\immediate
\write\rfile{\noexpand\item{#1\ }\reflabeL{#1\hskip.31in}\pctsign}\findarg}
\def\findarg#1#{\begingroup\obeylines\newlinechar=`\^^M\pass@rg}
{\obeylines\gdef\pass@rg#1{\writ@line\relax #1^^M\hbox{}^^M}%
\gdef\writ@line#1^^M{\expandafter\toks0\expandafter{\striprel@x #1}%
\edef\next{\the\toks0}\ifx\next\em@rk\let\next=\endgroup\else\ifx\next\empty%
\else\immediate\write\wfile{\the\toks0}\fi\let\next=\writ@line\fi\next\relax}}
\def\striprel@x#1{} \def\em@rk{\hbox{}} 
\def\lref{\begingroup\obeylines\lr@f}
\def\lr@f#1#2{\gdef#1{\defref#1{#2}}\endgroup\unskip}
\newwrite\lfile
{\escapechar-1\xdef\pctsign{\string\%}\xdef\leftbracket{\string\{}
\xdef\rightbracket{\string\}}}

\def\writestop{\def\writestoppt{\immediate\write\lfile{\string\p
ageno%
\the\pageno\string\startrefs\leftbracket\the\refno\rightbracket%
\string\def\string\secsym\leftbracket\secsym\rightbracket%
\string\secno\the\secno\string\meqno\the\meqno}\immediate\closeout\lfile}}
\def\writestoppt{}\def\writedef#1{}
\catcode`\@=12 

\rightline{DAMTP-2001-38}
\rightline{M/C-TH-01/03}
\vskip 9pt
\centerline{\fourteenbf New data and the hard pomeron}
\vskip 8pt
\centerline{A Donnachie}
\centerline{Department of Physics, Manchester University}
\vskip 5pt
\centerline{P V Landshoff}
\centerline{DAMTP, Cambridge University$^*$}
\footnote{}{$^*$ email addresses: ad@a35.ph.man.ac.uk, \ pvl@damtp.cam.ac.uk}
\bigskip
{\bf Abstract}

New structure-function data are in excellent agreement with the existence
of a hard pomeron, with intercept about 1.4.
It gives a very economical description of the data.
Having fixed 2
parameters from the data for the real-photon cross section 
$\sigma^{\gamma p}$, we need just 5 further parameters
to fit the data for $F_2(x,Q^2)$ with
$x\leq 0.001$. The available data range from $Q^2=0.045$ to 35 GeV$^2$.
With guesses consistent 
with dimensional counting for the $x$ dependences of our three separate terms, 
the fit extends well to larger $x$ and to
$Q^2=5000$ GeV$^2$. With no additional parameters, it gives a good description
of data for the charm structure function $F_2^c(x,Q^2)$ from $Q^2=0$
to 130 GeV$^2$.
The two pomerons also give a good description of both the $W$ and the $t$
dependence of $\gamma p\to J/\psi\,p$.

\vskip 20truemm

In previous papers, we have shown that the Regge approach provides a
very good description\defref\twopom{
A Donnachie and P V Landshoff, Physics Letters B437 (1998) 408
} 
of the data on the small-$x$ behaviour of the proton
structure function $F_2(x,Q^2)$, on\defref\charm{
A Donnachie and P V Landshoff, Physics Letters B470 (1999) 243
}
the charm structure function $F_2^c(x,Q^2)$, and on\defref\rrho{
A Donnachie and P V Landshoff, Physics Letters B478 (2000) 146
}
exclusive photoproduction of the $J/\psi$. These data call for a new
Regge trajectory, whose intercept is about 1.4, which we call the
hard pomeron. In conformity with traditional Regge theory\defref\collins{
P D B Collins, {\it Introduction to Regge Theory and High Energy Physics},
Cambridge University Press (1977)
}
the hard-pomeron has an intercept that
does not vary with $Q^2$, but its contribution
is added to that of the soft pomeron and, because their relative weight is
$Q^2$-dependent, their combined effect is similar to that of a single
$Q^2$-dependent trajectory. The conventional theory of perturbative
evolution appears rather to favour a genuinely
$Q^2$-dependent trajectory; however, we have
explained\defref\perturb{
J R Cudell, A Donnachie and P V Landshoff, Physics Letters B448 (1999) 281
}
that the way in which the evolution is usually applied is
not valid at small $x$. Resummation is known to be necessary and to
have a very significant effect, but at present we do not have adequate
knowledge to implement it.

Because of this, it is important to study data, to try and extract whatever
theoretical message they may contain. Fits with huge numbers of parameters
are unlikely to reveal such messages at all clearly, but the Regge approach
has the merit that it requires rather few parameters. There have
recently appeared new and more accurate data\defref\zeus{ 
ZEUS collaboration: J Breitweg et al, Physics Letters B487 (2000) 53
}
for $F_2(x,Q^2)$ at small $Q^2$ from ZEUS and\defref\hone{
H1 collaboration: C Adloff et al, hep-ex/0012052 and hep-ex/0012053
}
at larger values of $Q^2$ from H1, who have also finalised\defref\jpsi{
H1 collaboration: C Adloff et al, Physics Letters B483 (2000) 23
}
their data on exclusive $J/\psi$ photoproduction. 
In this paper we confront these new measurements with the Regge approach,
and find that it stands up well.

Our previous fit\ref{\twopom} to the data for $F_2(x,Q^2)$ used simple powers of
$x$:
$$
F_2(x,Q^2) = \sum_{i=0}^2f_i(Q^2)\,x^{-\epsilon_i}
\eqno(1)
$$
Here, the $i=0$ term is hard-pomeron exchange, $i=1$ is soft-pomeron
exchange, and $i=2$ is ($f_2,a_2$) exchange.
Our fit extended up
to $x=0.07$. However, the new data are so very accurate, with errors just a few
percent, that it is no
longer safe to go to such large $x$ with simple powers: they
will be modified by unknown factors which eventually
ensure that $F_2(x,Q^2)$ vanishes at each $Q^2$ when $x\to 1$. 
The dimensional-counting rules would require both the soft and the hard pomeron
contributions to behave near $x=1$ as $(1-x)^7$, 
and the ($f_2,a_2$) contribution as $(1-x)^3$, but it is not clear that
these rules are valid and there is no theoretical
information about how the factors should behave away from $x=1$.
Nevertheless, we make simple guesses, which conform with the 
dimensional-counting rules, and which  
are probably a better approximation than omitting the factors altogether:
$$
F_2(x,Q^2)=f_0(Q^2)\,x^{-\epsilon_0}(1-x)^7+f_1(Q^2)\,x^{-\epsilon_1}(1-x)^7
+f_2(Q^2)\,x^{-\epsilon_2}(1-x^2)^3
\eqno(2)
$$
We explain below our reason for using $(1-x^2)^3$ for the ($f_2,a_2$)
term, rather than simply $(1-x)^3$.
Our fit uses only data up to $x=0.001$, where the factors
have less than 1\% effect, but we will find
that we have quite good agreement with the data even beyond
$x=0.07$, where they are rather important.

As we have done before\ref{\twopom},
we fix the soft-pomeron power $\epsilon_1$ at the value 0.0808
which we found\defref\sigtot{
A Donnachie and P V Landshoff, Physics Letters B296 (1992) 227
}
from data for $\sigma^{pp}$ and $\sigma^{\bar pp}$, though it has been
argued\defref\cudell{
J R Cudell et al, Physical Review D61 (2000) 034019; Erratum-{\it ibid} D63 (2001) 059901
}
that a rather larger value should be taken, perhaps 0.093.
There is little theoretical understanding of the functions $f_1(Q^2)$ in (1).
All we know is that near $Q^2 = 0$
$$
f_i(Q^2) \sim X_i \,(Q^2)^{1+\epsilon_i}
\eqno(3)
$$
To guide us on the likely functional forms of the coefficient
functions we extract their values at each $Q^2$ 
for which there are new data\ref{\zeus,\hone}
for $F_2(x,Q^2)$. In order to
have enough data points at each $Q^2$ we include values of $x$ up to
0.02, rather than the up to 0.001 as we have suggested above. Even up to 
$x=0.02$ the contribution from the ($f_2,a_2$) term is small, so we
omit it at this stage; that is, we use (2) without the last term. Figure 1
shows the coefficient functions $f_0(Q^2)$ and 
$f_1(Q^2)$ corresponding to two choices of
the hard-pomeron power, $\epsilon_0 = 0.36$ and $\epsilon_0 = 0.50$, which
lie either side of the value $\epsilon_0 = 0.44$ which we have recently
said\ref{\charm} is preferred. We stress that these fits should not be taken too
seriously: we use them only as a guide and then  go back to the beginning
with the fitting.

\topinsert
\line{\epsfxsize=0.45\hsize\epsfbox[70 590 300 770]{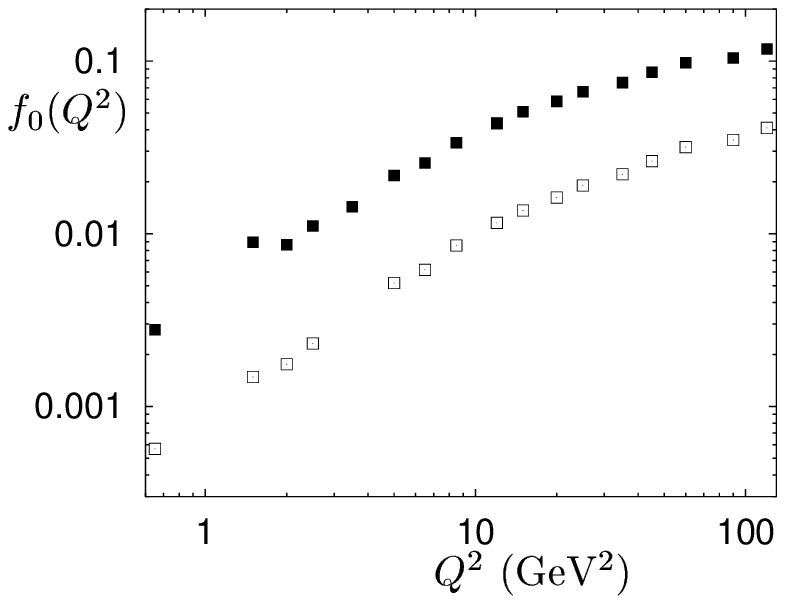}\hfill
\epsfxsize=0.45\hsize\epsfbox[70 590 300 770]{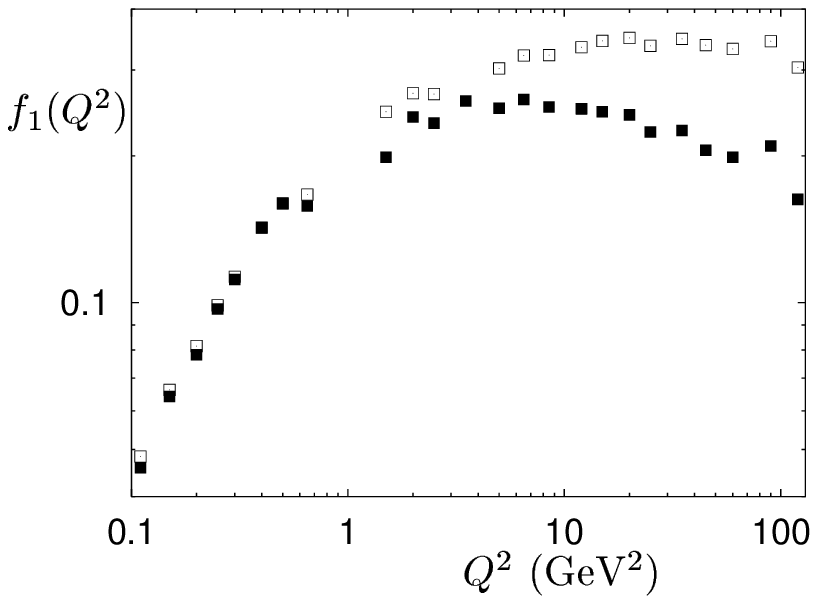}}
\line{\hfill (a)\hfill\hfill (b)\hfill}
Figure 1: The coefficient functions $f_i(Q^2)$ of (1) extracted from
the new data\ref{\zeus,\hone}, (a) for the hard pomeron and (b) for the soft.
The points are for $\epsilon_0=0.36$ (black points) and 0.5 (open points). 
\endinsert

By making a rough fit to the points in
figure 1a we may deduce that the shape of the hard-pomeron coefficient
$f_0(Q^2)$ is likely to be well-described by the form
$$
f_0(Q^2)=X_0\Big ({Q^2\over 1+Q^2/Q_0^2}\Big )^{1+\epsilon_0}
(1+Q^2/Q_0^2)^{\epsilon_0/2}
\eqno(4a)
$$
We introduced this form in our fit\ref{\charm} to the data on the
charm structure function $F_2^c(x,Q^2)$. It is more economical than
the shape we used originally\ref{\twopom}, in that it includes one
fewer parameter.
Figure 1b shows that the shape of the soft-pomeron coefficient
function $f_1(Q^2)$ is sensitive to the value for the hard-pomeron power
$\epsilon_0$. Our previous fits suggested that $f_1(Q^2)$ goes to
zero at large $Q^2$, 
but we find now that we obtain a good fit with a form that contains one
fewer parameter and obeys Bjorken scaling at large $Q^2$:
$$
f_1(Q^2)=X_1\Big ({Q^2\over 1+Q^2/Q_1^2}\Big )^{1+\epsilon_1}
\eqno(4b)
$$

Our fit uses data for $F_2(x,Q^2)$ with $x\leq 0.001$, where the contribution
from $f_2$ and $a_2$ exchange is small, less than 5\% according
to our results. However, 
we include also data for the real-photon total cross section
$\sigma ^{\gamma p}$, the very accurate pre-HERA data with
$6< W< 18$ GeV. For these data 
$f_2$ and $a_2$ exchange is important. We use the value we have
previously\ref{\sigtot}
taken for the ($f_2,a_2$) intercept,
$1+\epsilon_2$ with $\epsilon_2=-0.4525$. 
For want of adequate information, 
we assume that the corresponding coefficient function is similar in form to
(4):
$$
f_2(Q^2)=X_2\Big ({Q^2\over 1+Q^2/Q_2^2}\Big )^{1+\epsilon_2}
\eqno(4c)
$$
Then
$$
\sigma ^{\gamma p}(\nu)={4\pi ^2\alpha\over Q^2}F_2(Q^2/2\nu,\,Q^2)\Big\arrowvert
_{Q^2=0}=4\pi ^2\alpha \sum_{i=0}^2 X_i\, (2\nu)^{\epsilon_i}
\eqno(5)
$$

\topinsert
\line{\hfill\epsfxsize=0.40\hsize\epsfbox[90 430 290 760]{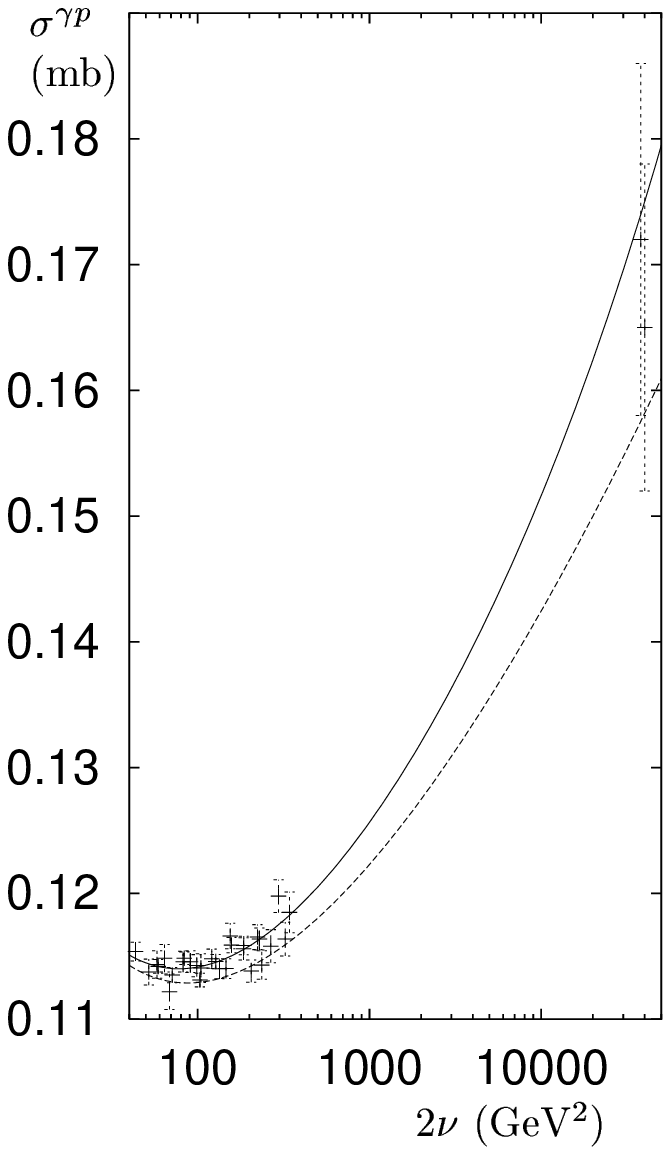}\hfill
\epsfxsize=0.40\hsize\epsfbox[90 430 290 760]{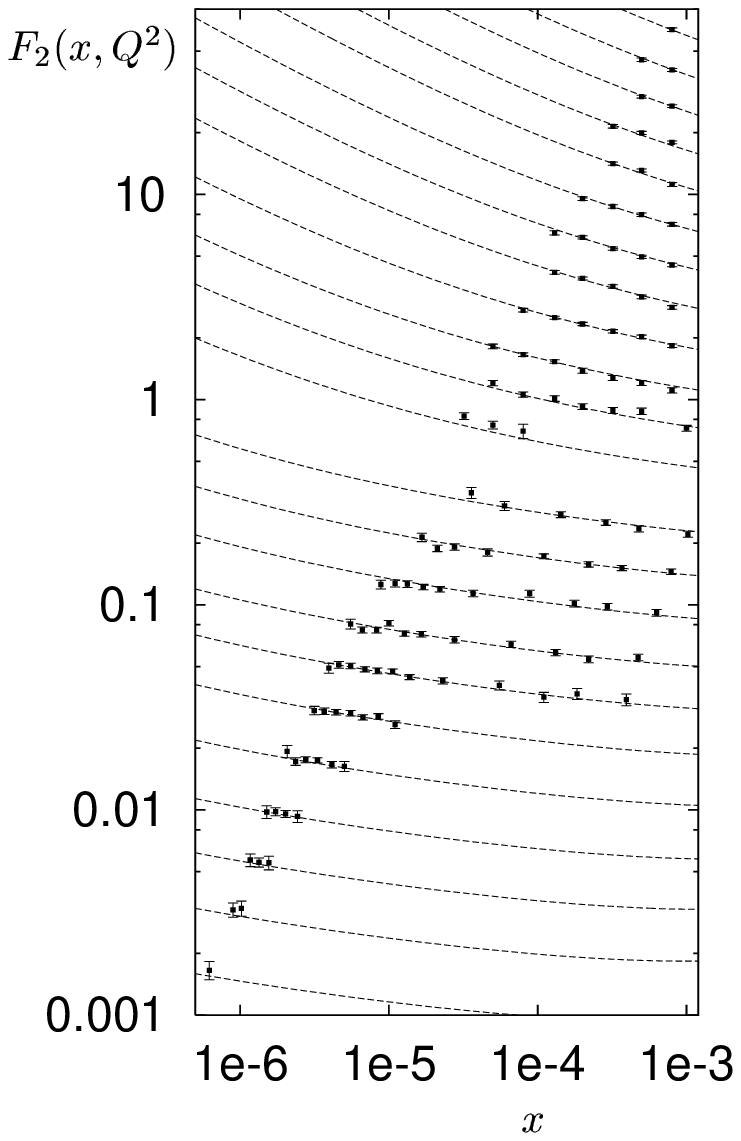}\hfill}
\vskip 3pt
\line{\hfill (a)\hfill\hfill (b)\hfill}
Figure 2: The fit described in the text to (a) the real-photon cross section
$\sigma^{\gamma p}$ and (b) to recent ZEUS and H1 data\ref{\zeus ,\hone}
for $F_2(x,Q^2)$. In (a) the lower line omits the contribution from
hard-pomeron exchange. In (b), the data for the different values of  $Q^2$ 
have been scaled with powers of $\sqrt 2$; $Q^2$
varies from 0.045 GeV$^2$ at the bottom to 35 GeV$^2$ at the top.
\endinsert

We use (2), (4) and (5), with  $\epsilon_0, X_0,
Q_0^2, X_1, Q_1^2, X_2$ and $Q_2^2$ as free parameters, to perform a 
least-squares fit to the new 
ZEUS and H1 data for $F_2(x,Q^2)$ up to $x=0.001$, together with the data for
$\sigma ^{\gamma p}$. The best values are 
$$
\epsilon_0=0.4372~~~X_0=0.001461~~~  Q_0^2= 9.108~~~  X_1=0.5954~~~  Q_1^2=0.5894~~~  X_2= 1.154~~~ Q_2^2=0.2305
\eqno(6)
$$
and the average $\chi^2$ for each of the 148 data points is 0.98.
In making our fit, we combine the statistical and 
systematic errors
in quadrature. In the case of the new H1 data, we ignore the correlated
errors. Figure 2 shows the data we have used, together with the fits.
\bigskip\bigskip
We make a number of comments on these fits:

{\bf 1}  Because the errors on the data are so small, the $\chi^2$
per data point is very sensitive to the precise values of the parameters~(6).
However, it should not be thought that
the parameters are determined to anything like the
quoted accuracy: one can change
any of them, with compensating changes to the others, and still achieve a
good  $\chi^2$. Furthermore, by completely ignoring the correlated errors
in the H1 data we have been much too conservative. So the error on
the value of $\epsilon_0$ is at least 10\%.

{\bf 2}  The real-photon cross section $\sigma^{\gamma p}$ plays an important
role in the fit and determines the values of the parameters
$X_1$ and $X_2$.  Without these data, $\epsilon_0$ could vary over a large
range, from less than 0.25 to more than 0.5, with the appropriate change of
shape for the soft-pomeron coefficient function $f_1(Q^2)$ suggested by
figure 1a, and still give a good $\chi^2$. The values of the $X_i$ in
(6) correspond to
$$
\sigma^{\gamma p}=0.00016\, (2\nu)^{\epsilon_0}+0.067\,(2\nu)^{\epsilon_1}+
0.129\,(2\nu)^{\epsilon_2}
\eqno(7)
$$
in mb units with $\nu$ in GeV$^2$. The values 0.067 and 0.129 for the
soft-pomeron and  $f_2,a_2$ coefficients are the same as in our original
fit\ref{\sigtot} without a hard-pomeron term. This is because,
as is evident from figure 2a, where the lower curve omits the hard-pomeron term
in (7), the hard-pomeron contribution
is small at $Q^2=0$ in the energy range of
the pre-HERA data, smaller than we had it previously\ref{\twopom}
but not completely negligible. We have explained before\ref{\perturb}
that the question whether the hard-pomeron term is present already at
$Q^2=0$ is of considerable theoretical importance. If it is present then
the hard pomeron is not, as many people think, generated by 
perturbative evolution, though the observed increase with $Q^2$ of its
importance may be a consequence of perturbative evolution.

{\bf 3} We have previously noted\ref{\charm}
that the charm structure function $F_2^c(x,Q^2)$ is well described by
hard-pomeron exchange alone, and that hard-pomeron exchange
seems to be flavour blind even at low $Q^2$. 
Figure 3a shows ZEUS data\defref\zeusc{
ZEUS collaboration: J Breitweg et al, European Physical Journal C12 (2000) 35
}
together with
${2\over 5}$ of the hard-pomeron contribution to the complete
$F_2(x,Q^2)$ used in figure 2b. The fraction ${2\over 5}$ is
$e^2_c/(e^2_u+e^2_d+e^2_s+e^2_c)$. This zero-parameter fit works well all
the way from $Q^2=1.8$ to 130 GeV$^2$. It even continues to be satisfactory
when extrapolated to $Q^2=0$: see figure~3b.
\topinsert
\centerline{\epsfysize=0.45\vsize\epsfbox[90 430 370 760]{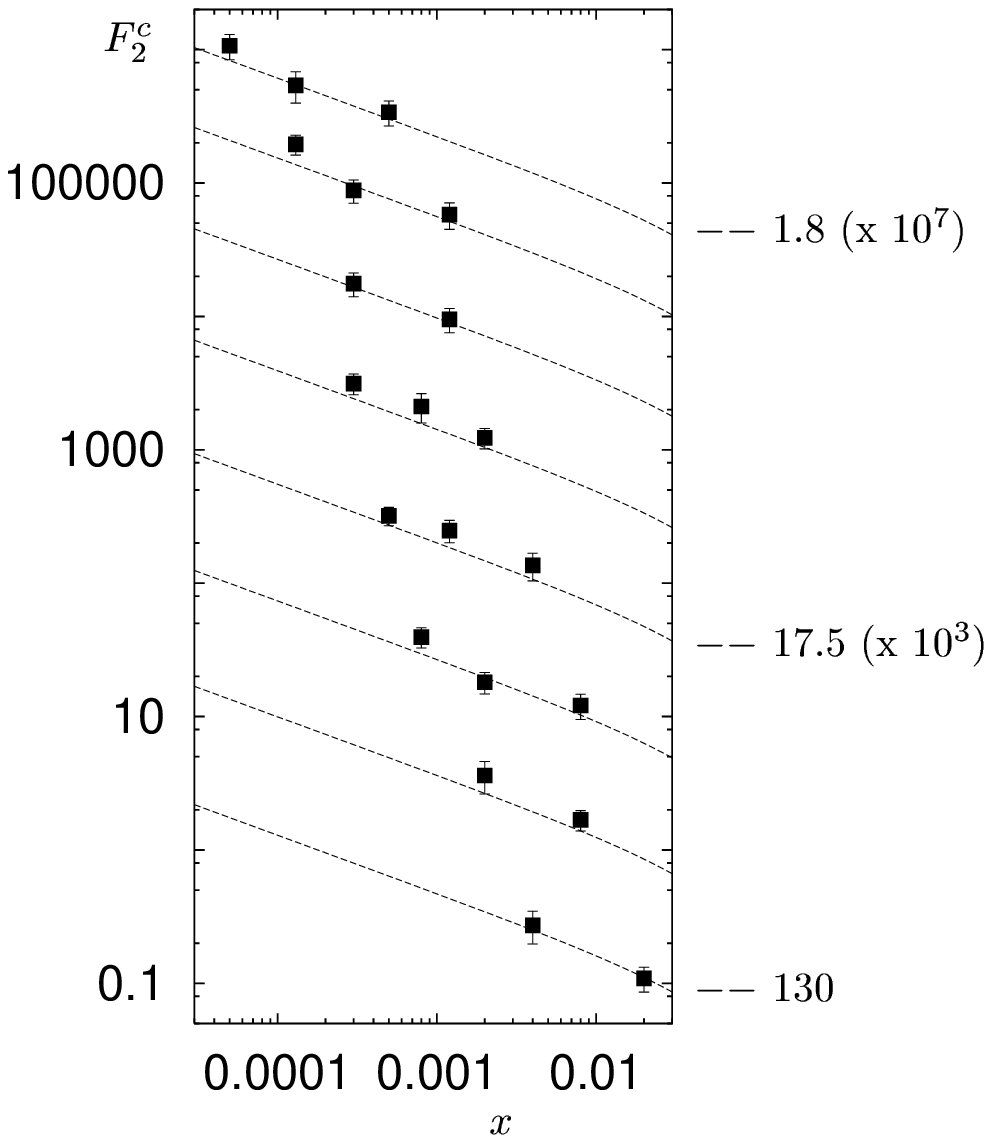}\hfill
\epsfxsize=0.4\hsize\epsfbox[80 600 300 770]{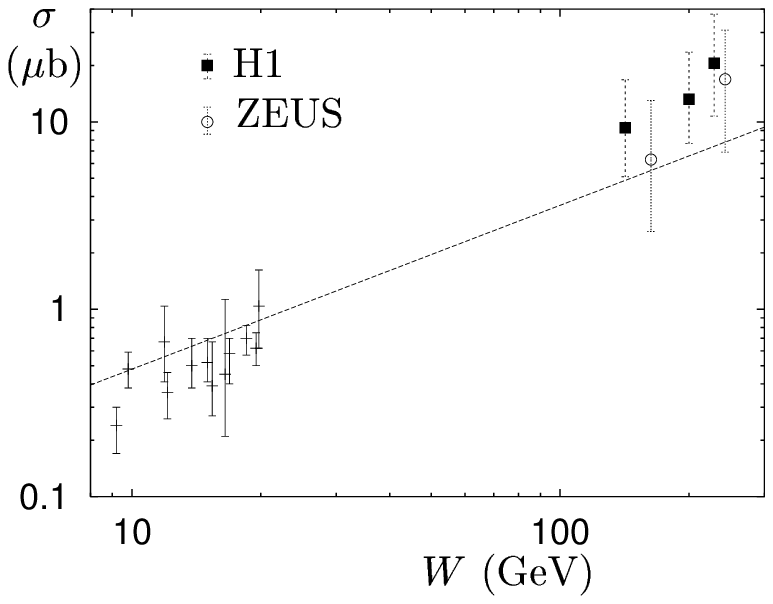}}\h
\line{\hfill (a)\hfill\hfill (b)\hfill}

Figure 3: (a) ZEUS data\ref{\zeusc} for the charm structure function 
with the zero-parameter hard-pomeron fit. The data for the various values of
$Q^2$ have been scaled with powers of 10, with $Q^2=1.8$ GeV$^2$ at the top
and 130 GeV$^2$ at the bottom. (b) data for the charm photoproduction 
cross section; the fit is that shown in (a), extrapolated to $Q^2=0$.
\endinsert

{\bf 4} We may check that the values of $X_2$ and $Q_2^2$ in (6) are
consistent with data for real and virtual photons scattering on a
neutron target. These parameters relate to the sum of $f_2$ and
$a_2$ exchange. When we switch to a neutron target, the contribution
from $a_2$ exchange changes sign.  So the difference $F_2^p(x,Q^2)-
F_2^n(x,Q^2)$, which has been measured by the EMC collaboration\defref\emc{
J J Aubert et al, Nuclear Physics B293 (1987) 740
},
corresponds just to $a_2$ exchange. We assume that the $Q^2$ dependence
of $f_2$ and $a_2$ exchanges have the same shape, that is both are given by
(4c) with the same value for $Q_2^2$, given in (6). 
The $x$ dependence of the data is well described by including a
factor $(1-x^2)^3$, which is what led us to the choice of the last term
in (2). With such a factor,
we fit the ($p-n$) data with a $\chi^2$ of 0.82 per data point, with
$$
X^{p-n}_2=0.37
\eqno(8)
$$
This corresponds to $a_2$ exchange having about $1\over 5$ the strength
of $f_2$ exchange. The $Q^2$ values of the EMC data vary between 7 and
170 GeV$^2$. We assume that this factor of $1\over 5$ remains valid
down to $Q^2=0$. In figure 4a the fit (7) is extrapolated to low energies and is
compared with the data for
$\sigma^{\gamma p}$. In figure 4b the data are for $\sigma^{\gamma n}$
and the upper curve is the same fit, while the lower curve has the coefficient
of the last term multiplied by ${2\over 3}$.   This lower value corresponds
to the contribution from $a_2$ exchange to $\sigma^{\gamma p}$
being $1\over 5$ that of $f_2$
exchange. It gives a reasonable eyeball fit and 
it provides a consistency check.

\topinsert
\line{\epsfxsize=0.45\hsize\epsfbox[85 590 290 770]{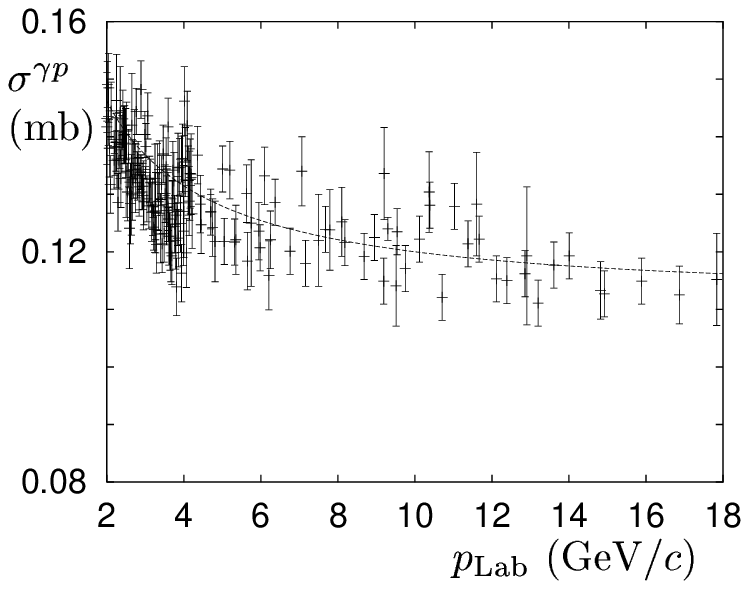}\hfill
\epsfxsize=0.45\hsize\epsfbox[85 590 290 770]{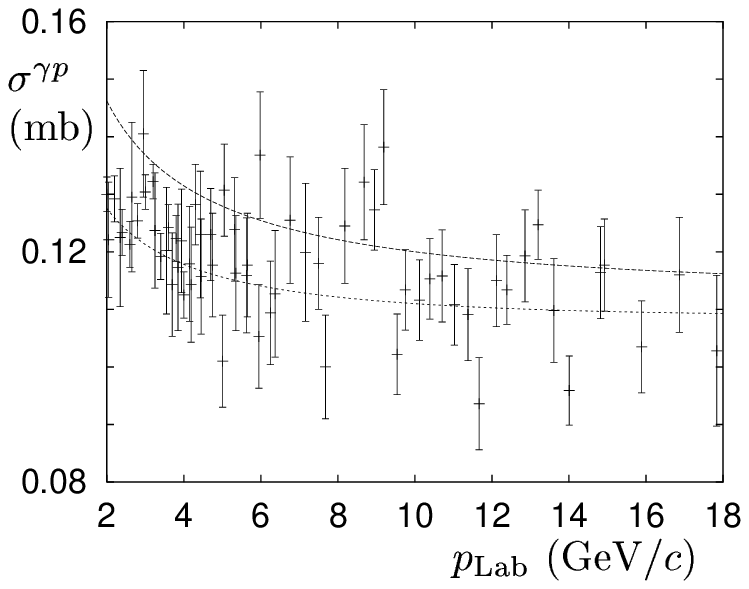}}
\line{\hfill (a)\hfill\hfill (b)\hfill}
Figure 4: Data for (a) $\sigma^{\gamma p}$ and (b) $\sigma^{\gamma n}$.
The curve in (a) and the upper curve in (b) is the fit (7); the lower curve
in (b) has the last term reduced by a factor ${2\over 3}$.
\endinsert
 
{\bf 5} We have previously shown that a combination of soft and hard pomeron
exchange gives a good description of the differential cross section
for the exclusive process $\gamma~p\to J/\psi~p$. The preliminary data on
which we based our fit\ref{\charm} have now been finalised\ref{\jpsi}.
We fit the data 
assuming that the amplitude is 
$$
A(s,t)=i\sum_{i=0,1}\beta_i(t)\,(\alpha'_i s)^{e_i(t)}
e^{-{1\over 2}i\pi e_i(t)}
\eqno(9)
$$
Here, $1+e_i(t)$ are the two pomeron trajectories, so that
$e_i(0)=\epsilon_i$. We assume that both are linear, and call
their slopes $\alpha'_i$. Before we raise $s$ to the power $e_i(t)$
we need to divide it by a squared-mass scale to make it dimensionless. 
We decided long ago\defref\alphaprime{
A Donnachie and P V Landshoff, Nuclear Physics B267 (1986)  690
}
that it was appropriate to use $1/\alpha'_i$ for each power. We found 
also\ref{\charm},
when considering the preliminary data from H1, 
that the data were well described by assuming that the coupling functions
are
$$
\beta_i(t)=b_{i}\,F_1(t)
\eqno(10)
$$
with $b_{i}$ constants and $F_1(t)$ the Dirac elastic form factor of the
proton:
$$
F_1(t)={4m^2-2.79t\over 4m^2 -t}{1\over (1-t/0.71)^2}
\eqno(11)
$$
where $m$ is the proton mass. It is known\ref{\alphaprime}
that the coupling
of the soft pomeron to the proton varies as $F_1(t)$ and it is reasonable to
assume that the same is true for the hard pomeron. So in using (11) we are
assuming that each pomeron has constant coupling to the $\gamma$-$J/\psi$
vertex.  The phases $e^{-{1\over 2}i\pi e_i(t)}$ are the standard
Regge phases.
The value of the soft-pomeron slope is well established\ref{\alphaprime}
to be 0.25~GeV$^{-2}$. For the hard-pomeron slope we take the same value
as we have used before, 
$$
\alpha'_0=0.1 \hbox{ GeV}^{-2}
\eqno(12)
$$
We integrate $d\sigma/dt$ calculated from (9) over $t$ and fit
to the total cross section for $\gamma p\to J/\psi\,p$,
using the new H1 data and fixed-target data. The best fit gives
$$
b_0=0.46~~~~~~~~~~b_1=5.4
\eqno(13)
$$
where the units are such that $|A(s,t)|^2$ is the differential cross section
$d\sigma/dt$ in nb GeV$^{-2}$ units: see figure 5a. The data for  
$d\sigma/dt$ are given in bins of $t$; we average the $|A(s,t)|^2$ over
each bin, rather than simply evaluating it at the average value of $t$
for the bin. The result is shown in figure 5b. 
Notice that, although neither pomeron has zero slope, 
the combined effect of the two pomerons has
almost no shrinkage: the curves in figure 5b are rather parallel. 
One can understand how this comes about, because of our choice
$\alpha'_0=0.1$. 
It is obvious that choosing the value $\alpha'_0 = \alpha'_1=0.25$ would
lead to shrinkage, while going to the limit $\alpha'_0 = 0$ would
produce negative shrinkage because
as the energy increases the relatively-steep soft-pomeron term
gives way to the flat hard-pomeron term.

\topinsert
\centerline{\epsfxsize=0.45\hsize\epsfbox[85 600 300 770]{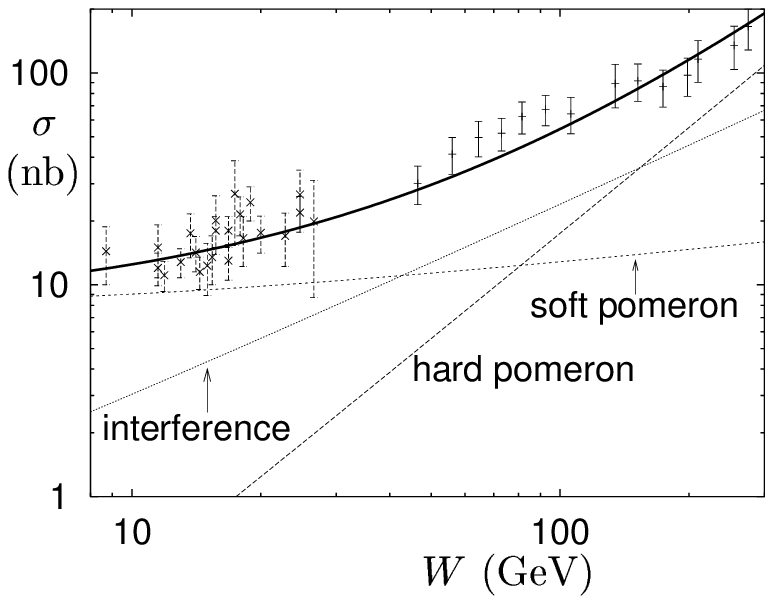}}
\vskip 2pt
\centerline{(a)}
\vskip 5mm
\hskip 30truemm{\epsfxsize=0.55\hsize\epsfbox[50 600 300 770]{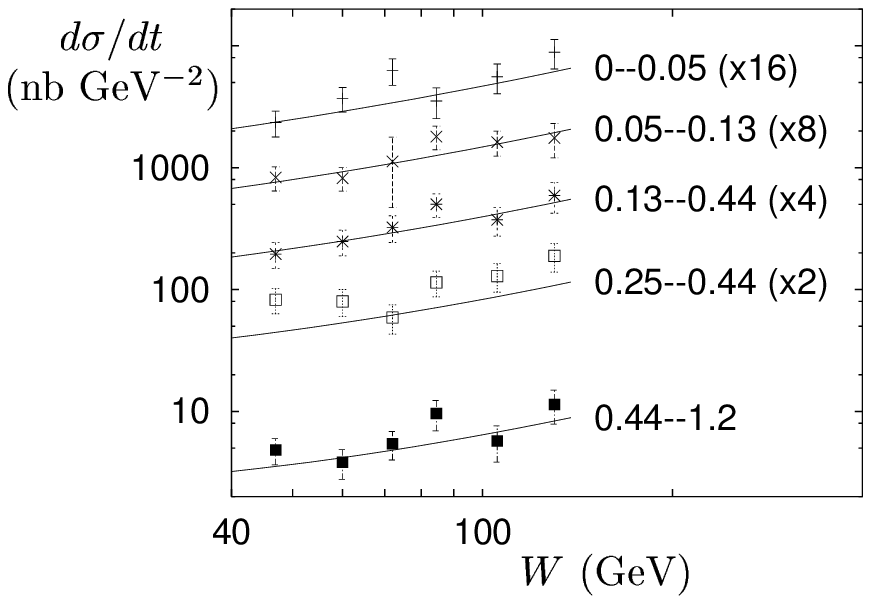}}
\vskip 2pt
\centerline{(b)}

Figure 5: H1 data\ref{\jpsi} for exclusive $J/\psi$ production; (a) the
total cross section and (b) the differential cross section for various
ranges of $t$.
The fits are described in the text. 
\endinsert

{\bf 6} Regge theory should be applicable to larger values of $x$ than 0.001,
provided sufficient nonleading exchanges are included. For want of any proper
information about these, and to avoid introducing additional parameters,
we have already said that we
take them into account by including in each of the two pomeron terms
a factor $(1-x)^7$, and in the ($f_2,a_2$)-exchange term a factor 
$(1-x^2)^3$. These powers agree with what the dimensional-counting rule
would require for the behaviour as $x\to 1$; we have no right to expect that
such simple behaviour is correct away from $x=1$, but it does work
surprisingly well. In the Regge framework, the nonleading powers of $x$
that we obtain when we multiply out these powers of $(1-x)$ are interpreted
as corresponding to the exchange of daughter trajectories.
The concept of daughter trajectories was developed\defref\daughter{
D Z Freedman and J M Wang, Physical Review 153 (1967) 1596
}
at a time when little was known about meson spectroscopy, in order
to cancel unwanted singularities in two-body nonelastic 
amplitudes such as $\pi^-\pi^+ \to K^-K^+$. This cancellation needs there
to be an infinite sequence of daughter trajectories corresponding to
each parent trajectory, with intercepts spaced one unit apart.
If, as seems to be a good approximation, trajectories are straight, 
this implies that the existence of a particle of spin $J$ implies
that there also exist particles of spin $J-1,\,J-2,\dots$ with the same
mass and the same other quantum numbers as the parent particle.
The data tables reveal the extensive existence of candidates
for daughter particles; figure 6 shows the $f$ family as an example.
We assume that the two pomerons have daughters too. The theory tells us
nothing about the $Q^2$ dependence of the daughter contributions; our assumption
that it is exactly the same as that of their parents is made to avoid
having to introduce additional parameters and seems to work well.
\topinsert
\centerline{\epsfxsize 0.5\hsize\epsfbox[50 600 300 770]{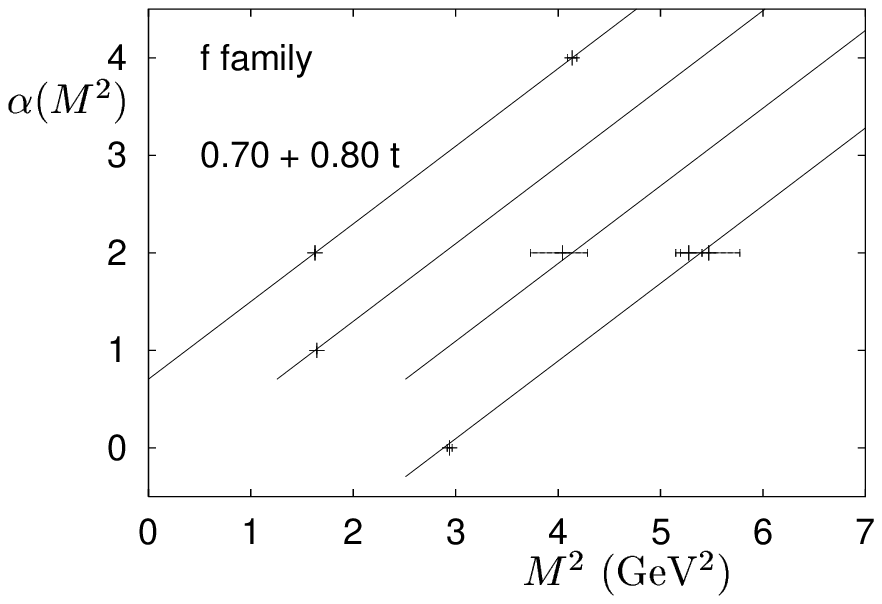}}
Figure 6: The $f$  family of Regge trajectories
\endinsert
\pageinsert
\centerline{\epsfysize 0.8\vsize\epsfbox[100 300 450 770]{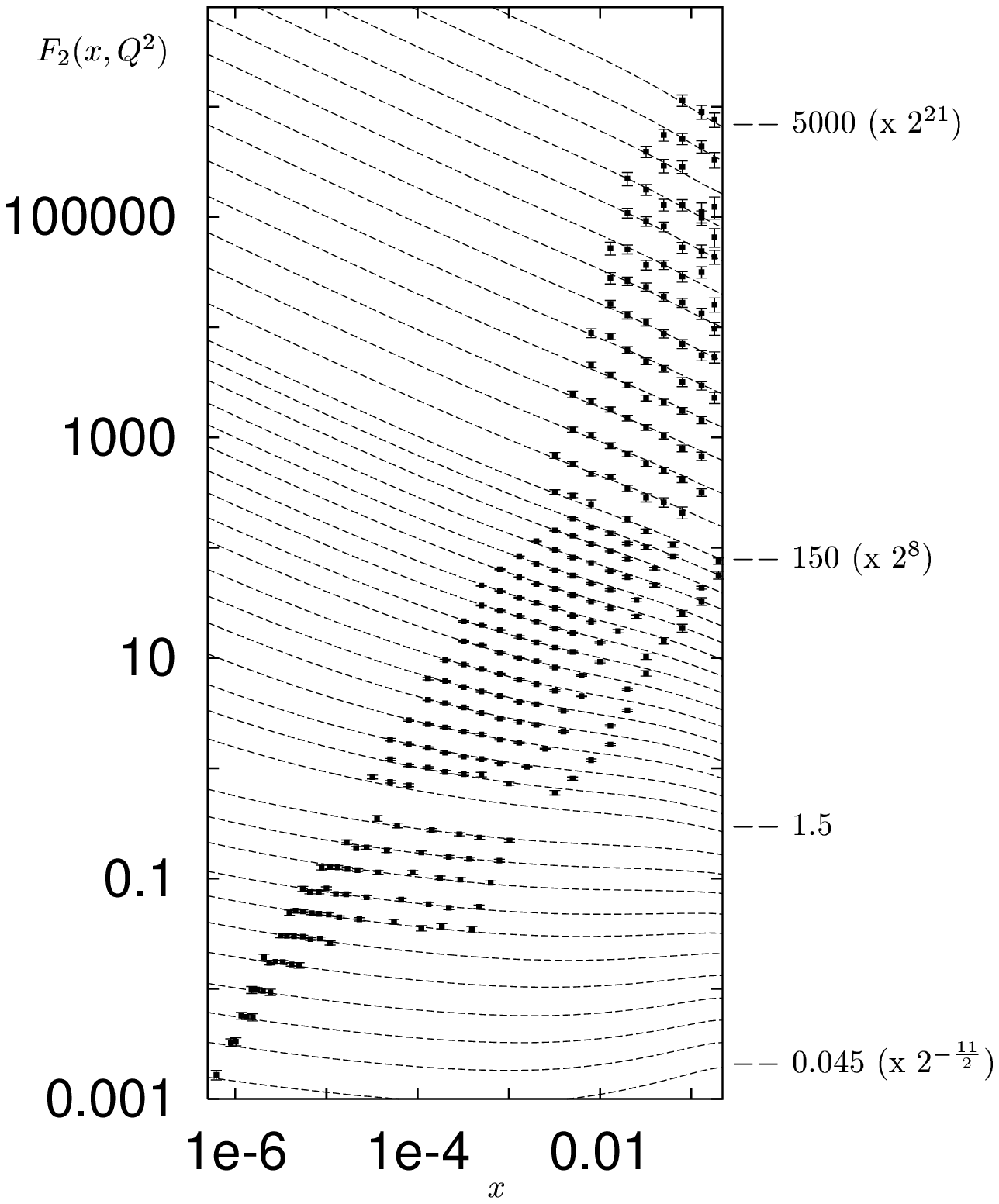}}
Figure 7: The fit described in the text compared with ZEUS and H1 
data\ref{\zeus,\hone}. The data for the various $Q^2$ values have been 
scaled with powers of $\sqrt 2$, from $Q^2=0.045$ GeV$2$ at the bottom
to  5000 GeV$2$ at the top.
\endinsert

{\bf 7} We saw in figure 1b that, according to the data, the shape of
the soft-pomeron coefficient function $f_1(Q^2)$ is sensitive to the value
of the hard-pomeron power $\epsilon_0$. The value $\epsilon_0=0.44$
gives a good fit to the data for $x<0.001$ with an $f_1(Q^2)$
given in (4b), which
goes to a constant at large $Q^2$. In our previous 
fit\ref{\twopom} we included also a factor $(1+Q/\bar Q)^{-1}$ in (4b). If we
do so now, and arbitrarily choose $\bar Q=10$ GeV, then the $\chi^2$
per data point for the fit to the real-photon plus $x<0.001$ data changes
only imperceptibly, and is still less than 1. It corresponds to the parameter
values
$$
\epsilon_0=0.3936~~~X_0=0.002475~~~  Q_0^2= 10.04~~~  X_1=0.5928~~~  Q_1^2=0.6643~~~  X_2= 1.150~~~ Q_2^2=0.2603
\eqno(14)
$$
Choosing a yet smaller value for $\bar Q$, with correspondingly a smaller
value for $\epsilon_0$ still gives an acceptable fit. 

{\bf 8} It is interesting to see how the fits compare with data at values of
$x$ larger than 0.001. There is no reason why they should perform well,
because our simple choices of multiplicative factors $(1-x)^7$ in the pomeron
terms and $(1-x^2)^3$ in the ($f_2,a_2$) term are surely too simple. Figure~7
shows the comparison with the data, for the parameter set (14).
We emphasise that our fit used only data for $x\leq 0.001$ and, as the
parameters $X_1$ and $X_2$ were determined from the data for 
$\sigma^{\gamma p}$, only 5 parameters were adjusted to fit $F_2(x,Q^2)$.

{\bf 9} We have shown that the Regge approach to the data for the structure
function $F_2(x,Q^2)$ gives a very simple and remarkably successful
description. There remains the urgent need to reconcile this approach
with perturbative QCD, but that cannot be done so long as there is so
little understanding of the theory of perturbative QCD at small $x$.

\topinsert
\centerline{\epsfxsize=0.5\hsize\epsfbox[50 430 290 650]{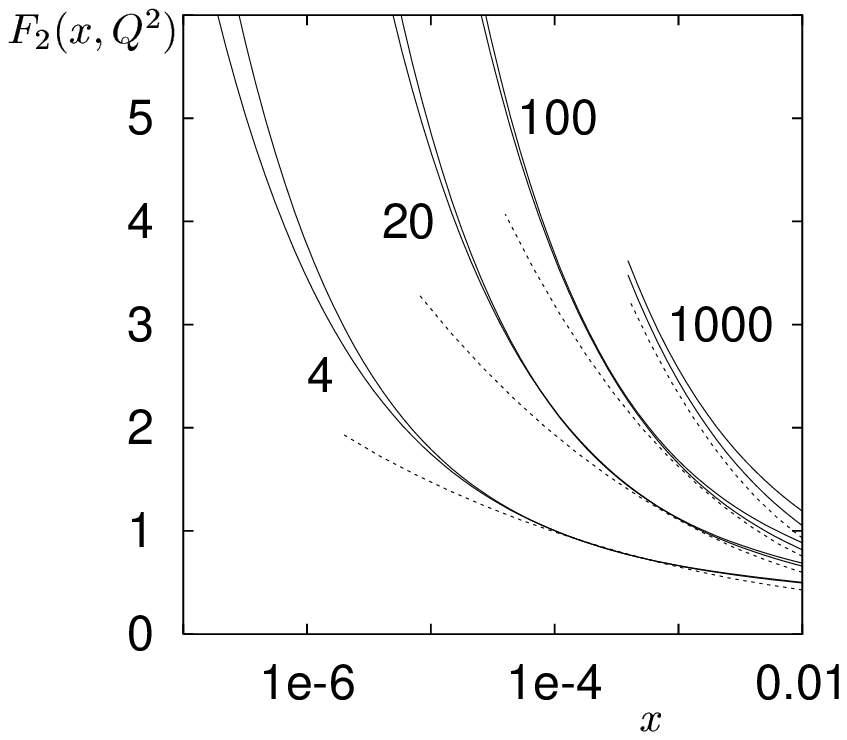}}
Figure 10:
The solid lines compare the fits with the parameter sets (6) (upper curves)
and (14) at various values of $Q^2$. 
The dashed curves are a two-loop pQCD fit\defref\abf{
G Altarelli, R D Ball and S Forte, hep-ph/0104246 }
\endinsert

{\bf 10} We have seen that the error on the extraction of the value of
the hard-pomeron intercept is quite large. Figure~10 illustrates this:
the pairs of solid curves are for the two parameter sets (6) and (14),
which have values of $\epsilon_0$ differing by 10\%. However, this figure
shows that it may be possible to use LHC and THERA data to distinguish
the Regge approach from the conventional DGLAP calculation with
the splitting function calculated to fixed order in $\alpha_s$. The
dashed curves show the result of such a fit\ref{\abf} using a two-loop
calculation\footnote{$^*$}{We are grateful to Richard Ball for making
available to us the numbers for the plot in his paper with Altarelli
and Forte\ref{\abf}}. Perhaps not surprisingly, the Regge approach predicts that
the rise at very small $x$ is more rapid. It is often said that unitarity
excludes such a rapid rise and forces it to be moderated by shadowing effects.
However, while shadowing should eventually set in, there is no reliable
way to estimate at what value of $x$ it will occur, and unitarity gives
no information. Unitarity places constraints of this type
only on purely-hadronic amplitudes: there is no
Froissart bound on $\sigma^{\gamma p}$ or $\sigma^{\gamma^* p}$,
because the unitarity equations are linear in the appropriate amplitudes.

\vfill\eject
\medskip\immediate\closeout\rfile\writestoppt
\baselineskip=14pt{{\bf References}}\bigskip{\frenchspacing%
\parindent=20pt\escapechar=` \input refs.tmp\bigskip}\nonfrenchspacing

\bye